\documentclass[prb,showpacs,twocolumn,preprintnumbers,amsmath,amssymb,floatfix]{revtex4}
\usepackage[dvips]{graphicx}
\usepackage[latin1]{inputenc}
\begin{document}
\draft

\newcommand{\lxpc} {Li$_{x}$ZnPc }
\newcommand{\lpc} {Li$_{0.5}$MnPc }
\newcommand{\etal} {{\it et al.} }
\newcommand{\ie} {{\it i.e.} }
\newcommand{\ip}{${\cal A}^2$ }

\hyphenation{a-long}

\title{Spin and charge dynamics in [TbPc$_2$]$^0$ and [DyPc$_2$]$^0$ single molecule magnets }

\author{F. Branzoli$^{1}$, P. Carretta$^{1}$, M. Filibian$^{1}$, M. J. Graf$^{2}$, S. Klyatskaya$^{3}$, M. Ruben$^{3,4}$, F. Coneri$^5$ and P. Dhakal$^{2}$.}

\address{$^{1}$Department of Physics ``A. Volta'', University of Pavia-CNISM, 27100 Pavia (Italy)}
\address{$^{2}$Department of Physics, Boston College, 02467 USA}
\address{$^{3}$Institute of Nanotechnology, Karlsruhe Institute of Technology (KIT), 76344 Eggenstein-Leopoldshafen (Germany)}
\address{$^{4}$IPCMS-CNRS UMR 7504, Université de Strasbourg, 67034 Strasbourg (France)}
\address{$^{5}$Department of Physics, University of Parma-CNISM, 43100 Parma (Italy)}

\widetext

\begin{abstract}
Magnetization, AC susceptibility and $\mu$SR measurements have been performed in neutral phthalocyaninato
lanthanide ([LnPc$_2]^0$) single molecule magnets in order to determine the low-energy levels structure and to
compare the low-frequency spin excitations probed by means of macroscopic techniques, such as AC susceptibility,
with the ones explored by means of techniques of microscopic character, such as $\mu$SR. Both techniques show a
high temperature thermally activated regime for the spin dynamics and a low temperature tunneling one. While in
the activated regime the correlation times for the spin fluctuations estimated by AC susceptibility and $\mu$SR
basically agree, clear discrepancies are found in the tunneling regime. In particular, $\mu$SR probes a faster
dynamics with respect to AC susceptibility. It is argued that the tunneling dynamics probed by $\mu$SR involves
fluctuations which do not yield a net change in the macroscopic magnetization probed by AC susceptibiliy. Finally
resistivity measurements in [TbPc$_2]^0$ crystals show a high temperature nearly metallic behaviour and a low
temperature activated behaviour.
\end{abstract}

 \maketitle

\narrowtext

\section{Introduction}
The trend towards ever-smaller electronic devices is driving electronics to its ultimate molecular-scale limit,
which allows for the exploitation of quantum effects. Single Molecule Magnets (SMM) are among the most promising
materials to be used in molecular spintronic devices \cite{Spintr} or as logic units in quantum computers
\cite{Leue}, since they combine the classical macroscale properties of bulk magnetic materials with the advantages
of nanoscale entities, such as quantum coherence. It has already been shown theoretically \cite{Leue} that
molecular magnets can be used to build efficient memory devices \cite{Grover} and, in particular, one single
molecule can serve as a storage unit of a dynamic random access memory device \cite{Svetlana,Vitali}. In addition,
SMM have been recognized as suitable materials which can be used as contrast agents for medical diagnostics. In
fact, these molecules comprise the chemical and magnetic structure of the gadolinium chelates with the
superparamagnetism of the SPIOs (Superparamagnetic Iron Oxides), materials which represent some of the most common
MRI contrast agents in current use \cite{MRI}.

LnPc$_{2}$-based compounds (Pc=C$_{32}$H$_{16}$N$_{8}$ is phthalocyanine, Ln a lanthanide ion) represent a new
class of SMM characterized by a doubly-degenerate ground state and large magnetic anisotropy which yields a
remarkably large barrier to spin reversal. In particular, [TbPc$_{2}$]$^-$ and [DyPc$_{2}$]$^-$ complexes are the
first mononuclear systems exhibiting a very slow relaxation of magnetization \cite{Ishikawa} and consequently a
characteristic correlation time ($\tau_c$) for the spin fluctuations reaching several $\mu$s at liquid nitrogen
temperature.  At variance with transition metal molecular magnets, in the LnPc$_2$ molecules the electronic levels
of the ground-state multiplet, with angular momentum $J$, are mainly split by the strong anisotropy of the crystal
field (CF) at the Ln$^{3+}$ ion. Accordingly the separation between the double degenerate ground state and the
first excited levels can reach several hundreds of kelvin, thus yielding unprecedently large $\tau_c$ at cryogenic
temperatures. Moreover, it is noticed that the magnetic properties of these lanthanide complexes are strongly
influenced by the chemical and structural modification of the ligands through which the CF potential can be varied
\cite{JACS}. The CF configuration can be modified from the very dilute limit in [LnPc$_{2}$]$^-$[TBA]$^+$N[TBA]Br,
with N+1$\gg 1$ the number tetrabutylammonia units, to the concentrated limit consisting solely of neutral
double-decker molecules [LnPc$_{2}$]$^0$. Recently it has been shown that one or two-electron oxidation of
[LnPc$_{2}$]$^-$ yields a compression of the cage around Ln$^{3+}$ ion and an increase of the CF splitting
\cite{Ishi2}$^,$\cite{Ishi3}$^,$\cite{Ishi4}.

By means of nuclear magnetic resonance (NMR) and muon spin relaxation ($\mu$SR), the correlation time for the spin
fluctuations in the neutral compounds [TbPc$_{2}$]$^0$ and [DyPc$_{2}$]$^0$ was found to be close to 0.1 ms at 50
K \cite{Neutri}, about two orders of magnitude larger than the one previously observed in the non-oxidized
lanthanide based single molecule magnets. In [TbPc$_{2}$]$^0$ two different regimes for the spin fluctuations have
been evidenced: a high temperature thermally activated regime involving spin fluctuations, driven by spin-phonon
coupling, across a barrier $\Delta \simeq$ 880 K separating the $|J = 6, m = \pm 6 \rangle $ ground states and the
$|J = 6, m = \pm 5 \rangle $ first excited states, and a low temperature regime involving quantum fluctuations
within the twofold degenerate ground-state. For [DyPc$_{2}$]$^0$ a high temperature thermally activated regime is
also found, however it cannot be explained in terms of a single spin-phonon coupling constant.

Here we report an analysis of the static uniform spin susceptibility ($\chi_S$)  for the [DyPc$_{2}$]$^0$ complex,
from which a determination of the low energy level structure of the Dy$^{3+}$ ion spin multiplet J=15/2 is made.
On the basis of these results for the CF level splitting it is possible to analyze the temperature (T) dependence
of muon spin-lattice relaxation rate in the presence of an applied field, derived from previous $\mu$SR
measurements performed at the ISIS facility\cite{Neutri} and from new measurements performed at the Paul Scherrer
Institute (PSI). Accordingly, the spin-phonon couplings driving the transitions among the CF levels have been
derived and the characteristic correlation time for the spin fluctuations estimated. $\mu$SR results are compared
to the ones obtained by AC susceptibility measurements in the same samples. It is found that both AC
susceptibility and $\mu$SR probe a high temperature activated and a low temperature tunneling regime for the spin
dynamics. However, while the values estimated for the correlation times in the high temperature activated regime
are quite close, at low temperature the tunneling rate probed by $\mu$SR is much faster than the one derived by AC
susceptibility. Finally, transport measurements in [TbPc$_{2}$]$^0$ crystals are presented.
\begin{figure}[h!]
\vspace{6.8cm} \includegraphics{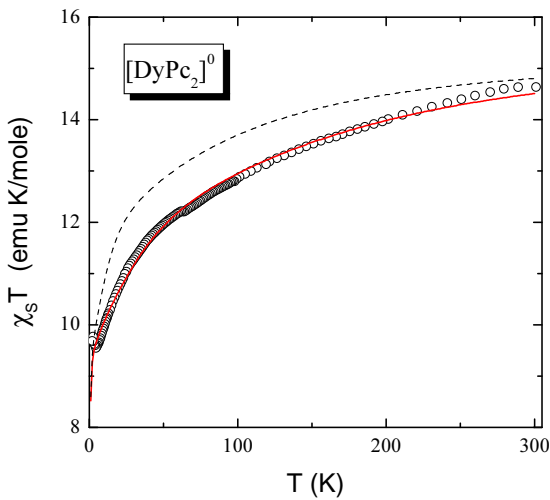}
\caption{\label{chiT}Temperature dependence of $\chi$T in [DyPc$_{2}$]$^0$ for H = 1000 Gauss (open circles) and
the best fit curve calculated according to Eq.1 curve (solid line). The dashed line corresponds to the behaviour
calculated on the basis of [DyPc$_{2}$]$^-$TBA$^+$ CF splittings derived by Ishikawa et al. \onlinecite{Ishi5}}
\end{figure}

\section{Technical aspects and Experimental Results}

All reagents were purchased from Across or Aldrich and used without further purification. The synthesis of both
samples was carried out by using several modifications of the protocol based on the published procedure reported
in Ref.\onlinecite{Sintesi}. The synthesis was accomplished by templating reactions, starting from a mixture of
the phthalonitrile precursor {\it o}-dicyanobenzene and the lanthanide acetylacetonate Ln($acac$)$_3\cdot$ {\it
n}H$_2$O, in the presence of a strong base (e.g. DBU, alkoxides) and high-boiling solvents, such as pentanol or
hexanol. A mixture of 1,2-dicyanobenzene (42 mmol), Ln(acac)$_3$ ·4H$_2$O (Ln = Tb$^{3+}$, Dy$^{3+}$) (5 mmol),
and 1,8- diazabicyclo[5,4,0]undec-7-ene (DBU) (21 mmol) in 50 mL of 1-pentanol was refluxed for 1.5 days. The
solution was allowed to cool to room temperature and then acetic acid was added and the mixture was heated at 100
C for 0.5 hours. The precipitate was collected by filtration and washed with  n-hexane and Et$_2$O. The crude
purple product was redissolved in 800 Ml of CHCl$_3$/MeOH (1/1) and the undissolved PcH$_2$ was filtered off.
 Both forms, blue (anionic [LnPc$_{2}$]$^-$) and green (neutral [LnPc$_{2}$]$^0$ ), were obtained
simultaneously, as revealed by electronic absorption spectra \cite{Sintesi}. In order to convert the unstabilized
anionic form to the neutral one, the reaction mixture was presorbed on active (H$_2$O-0\%) basic alumina oxide.
Purification was carried out by column chromatography on basic alumina oxide (deactivated with 4.6 \% H$_2$O,
level IV) with chlorophorm-methanol mixture (10:1) as eluent. In general, the yield was 30-35\%. By means of
additional radial chromatography on silica gel followed by recrystallization from chloroform-hexane mixture,
analytically pure powder samples were achieved.

\begin{figure}[h!]
\vspace{5 cm} \includegraphics{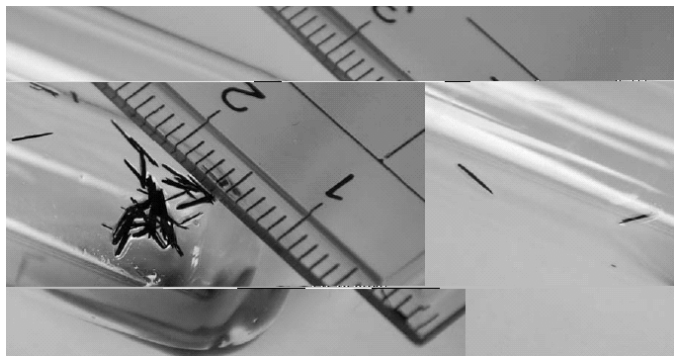} \caption{\label{crystal}
Photograph of a few [TbPc$_2$]$^0$ crystals grown according to the procedure reported above. The numbers in the
ruler scale correspond to centimeters.}
\end{figure}

Deep green crystals of the products were obtained by using slow diffusion of CH$_2$Cl$_2$ into C$_2$H$_2$Cl$_4$
solution of the pristine [LnPc$_2$]$^0$.  After 2 weeks, deep green needle-like crystals were obtained (see
Fig.\ref{crystal}). The [LnPc$_2$]$^0$ molecules crystallized in the space group $P2_12_12_1$ ($\gamma$-phase) as
reported in Ref.\onlinecite{Katoh} and were isomorphous to each other.

\begin{figure}[h!]
\includegraphics[width=18
pc]{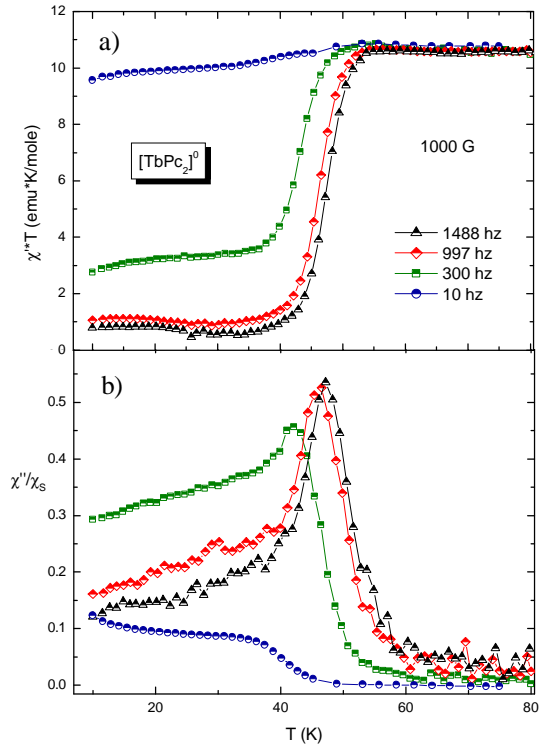}\hspace{0.6pc} \caption{\label{chiAcTb}Temperature dependence of a) $\chi'T$ and b)
$\chi''/\chi_S$ in [TbPc$_{2}$]$^0$, where $\chi'$, $\chi''$ and $\chi_S$ are in-phase-AC, out-of-phase-AC and DC
molar magnetic susceptibilities, respectively. The measurements were performed in 4 G oscillating magnetic field
at 10, 300, 997 and 1488 Hz, in the presence of a 1000 G DC component.}
\end{figure}

\begin{figure}[h!]
\vspace{7 cm} \includegraphics{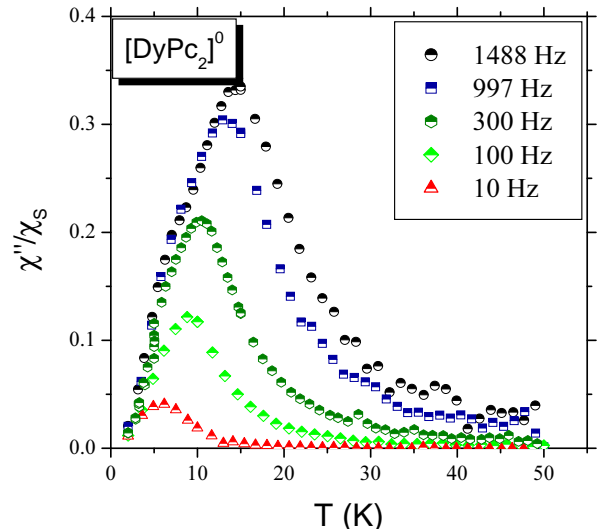}
\caption{\label{chiAcDy}Temperature dependence of the normalized out-of-phase spin susceptibility $\chi''/\chi_S$
in [DyPc$_{2}$]$^0$, for a 4 G oscillating magnetic field at various frequencies. A 9000 G static external field
was applied.}
\end{figure}

DC Magnetization (M) and AC susceptibility measurements have been performed by means of an MPMS-XL7 Quantum Design
SQUID magnetometer. The static uniform susceptibility was determined from the ratio $\chi_S= M/H$ in an applied
field $H= 1000$ G in the 2-300 K temperature range. In Fig. (\ref {chiT}) the T dependence of $\chi_S$T is shown.
In-phase ($\chi'$-real) and out-of-phase ($\chi''$-imaginary) AC susceptibility data were collected at
oscillating field frequencies between 10 and 1488 Hz, in the 10-80 K temperature range and for different applied
static fields. In Fig. (\ref{chiAcTb}) the temperature and frequency dependence of the AC susceptibility for
[TbPc$_{2}$]$^0$ is shown. A sharp drop in $\chi'T$ is observed on cooling in correspondence to a peak in
$\chi''/\chi_S$, which progressively shifts to lower temperatures as the AC frequency is lowered. In
[DyPc$_{2}$]$^0$ analogous but slightly broader peaks are observed at lower temperature (Fig. \ref{chiAcDy}),
suggesting a faster dynamic (see Discussion).

\begin{figure}[h]
\includegraphics[width=20
pc]{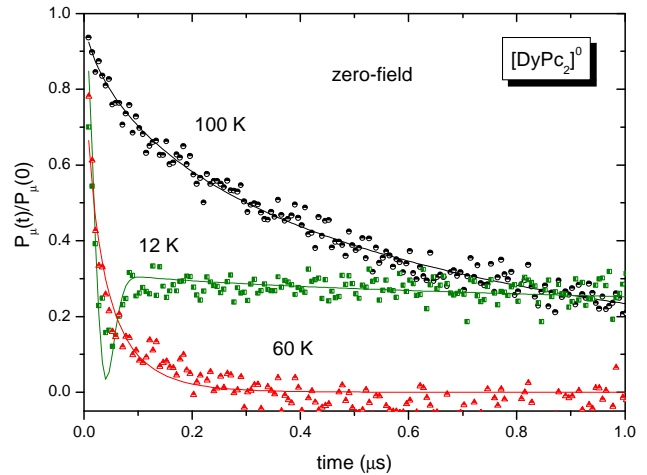}\hspace{0.6pc}%
\caption{\label{AsymmDy}Time evolution of the muon polarization in zero-field in [DyPc$_{2}$]$^0$ sample
normalized to its value for t $\rightarrow$ 0 at three selected temperatures. }
\end{figure}

Longitudinal field $\mu$SR measurements in the 50 mK-300 K temperature range were carried out either at ISIS
pulsed muon facility on MuSR beam line \cite{Neutri}, or at PSI muon facility on GPS (General Purpose Surface-Muon
Instrument) and LTF (Low Temperature Facility) beam lines. In contrast to the ISIS measurements, the continuous
beam muon production at PSI allows for a much more accurate determination of the short-time relaxation. In Fig.
(\ref{AsymmDy}) zero-field (ZF) depolarization curves at a few selected temperatures are shown for
[DyPc$_{2}$]$^0$. A Kubo-Toyabe (KT) relaxation\cite{Schenck} is clearly observed below $T^*\simeq 60$ K,
indicating very slow fluctuations, that is, $\nu /(\gamma_{\mu} \sqrt{\langle\Delta h^{2}_\bot\rangle_{\mu}})\ll
1$, with $\nu= 1/\tau_c$ being the characteristic frequency of the spin fluctuations and $\langle\Delta
h^{2}_\bot\rangle_{\mu}$ is the mean squared amplitude of the field fluctuations at the muon site. $\gamma_{\mu}$
is the muon gyromagnetic ratio. By fitting the low-temperature data with a KT function one can estimate the static
field distribution probed by the muons around $\sqrt{\langle\Delta h^{2}_\bot\rangle_{\mu}}$ = 490 G and
$\sqrt{\langle\Delta h^{2}_\bot\rangle_{\mu}}$ = 550 G for the Dy and Tb compounds, respectively. These values are
in excellent agreement with the measured "repolarization" of the signals with applied longitudinal fields. In the
slow fluctuations regime only the 1/3 tail of the KT depolarization function is affected by slow fluctuations and
it decays according to $P_{\mu}(t)= exp(-(2/3)\nu t)= exp(-\lambda t)$. On the other hand, for $T>T^*$ (around 60
K for [DyPc$_{2}$]$^0$ and around 90 K for [TbPc$_{2}$]$^0$) the fast fluctuations limit is attained.  In this
limit the decay of the polarization in zero or in a longitudinal field is given by $P_{\mu}(t) =
exp(-2\gamma_{\mu}^2 \langle\Delta h^{2}_\bot\rangle_{\mu}t/\nu )^\beta= exp(-\lambda t)^\beta$, where the
exponent $\beta$ can differ from unity in case of a distribution of muon sites, as it can be the case here. In the
intermediate fluctuation limit $\nu /(\gamma_{\mu} \sqrt{\langle\Delta h^{2}_\bot\rangle_{\mu}})\geq 1$
(T$\simeq$T*), $\lambda$ starts to decrease according to the so-called Abragam form\cite{Schenck}.

\begin{figure}[h!]
\vspace{6.8cm} \includegraphics{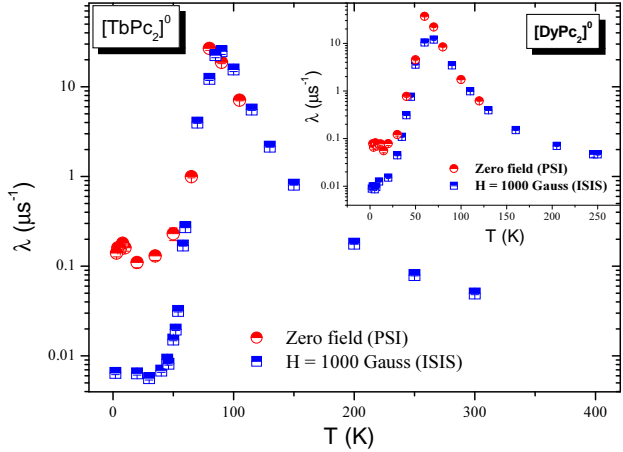} \caption{\label{lambdaT}T
dependence of the muon ZF longitudinal relaxation rate (circles) in [TbPc$_{2}$]$^0$ and [DyPc$_{2}$]$^0$ (in the
inset), together with 1000 G longitudinal field data from Ref. \onlinecite{Neutri} (squares).}
\end{figure}

\begin{figure}[h]
\includegraphics[width=20
pc]{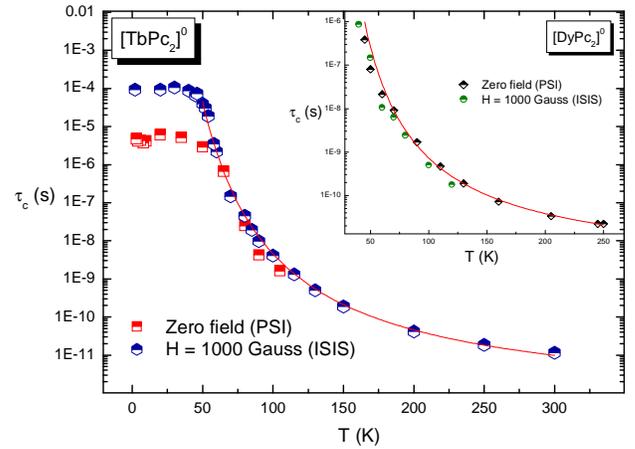}\hspace{0.8pc}%
\caption{\label{tau}T dependence of the correlation time for the spin fluctuations in [TbPc$_{2}$]$^0$ and
[DyPc$_{2}$]$^0$ (in the inset, for T $>$ T*) derived from $\lambda$ data reported in Fig. (\ref{lambdaT}). }
\end{figure}

By using the aforementioned expressions for the decay of the polarization in the fast and slow fluctuations regime
it was possible to derive the T-dependence of $\lambda$ (Fig.\ref{lambdaT}) in zero-field and estimate the
T-dependence of $\tau_c$ (see Fig. \ref{tau}), which was compared with the one previously derived at ISIS for a
1000 G longitudinal field \cite{Neutri}. The data for both materials were fit with a stretched exponential above
T$^*$,  with a T-independent exponent $\beta = 0.5$. In Fig. \ref{tau} one notices that in  the high T regime the
$\tau_c$ derived in zero-field and in a 1000 Gauss longitudinal field are in satisfactory agreement and show an
activated T-dependence of $\tau_c$. On the other hand, below 50 K, where tunneling processes become relevant,
$\tau_c(T)$ is observed to flatten at values which differ by more than an order of magnitude upon increasing the
field from zero to 1000 G. The increase of $\tau_c$ with the magnetic field intensity, also observed in the
AC-susceptibility measurements (Discussion), should be ascribed to the removal of the ground-state degeneracy by
the magnetic field, which progressively inhibits the tunneling processes.

\begin{figure}[h!]
\vspace{6.8cm} \includegraphics{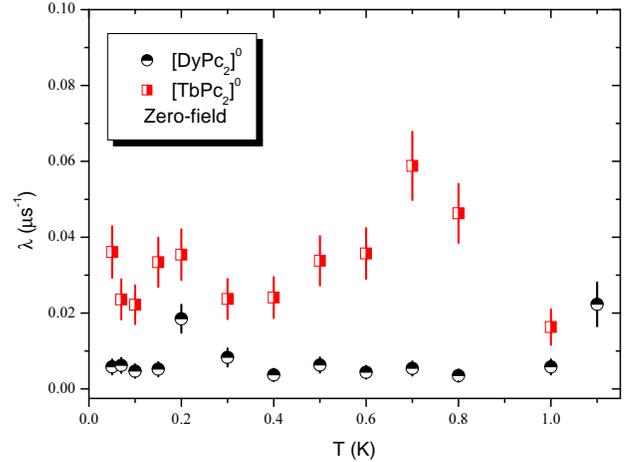} \caption{\label{LTFDyTb}T
dependence of the muon  relaxation rate in zero-field for [DyPc$_{2}$]$^0$ and [TbPc$_{2}$]$^0$, at T $<$ 1 K.}
\end{figure}

Data taken below 1 K in LTF yield an essentially temperature independent depolarization rate for both complexes
down to 50 mK, as expected for tunneling dominated depolarization. However, there exists a weak maximum in
$\lambda$(T), near T = 0.2 K for the Dy sample and near T = 0.7 K for the Tb sample, which warrant further studies
(Fig. (\ref{LTFDyTb}) and see Discussion).

\begin{figure}[h]
\vspace{2mm}
\includegraphics[width=20
pc]{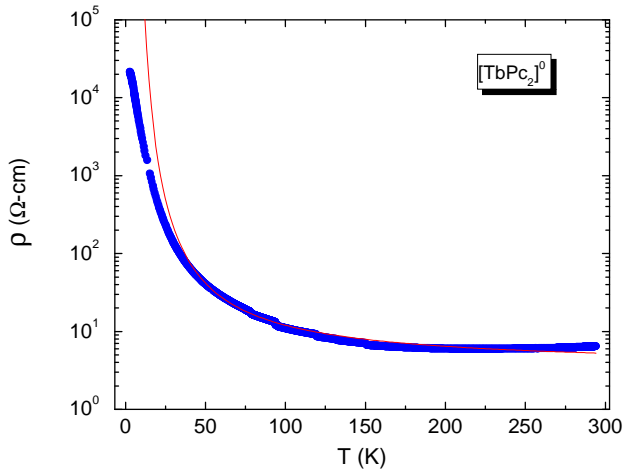}\hspace{0.6pc} \caption{\label{ro}Temperature
dependence of resistivity $\rho$, measured for a [TbPc$_{2}$]$^0$
crystal. Solid line represents the best fit of the activated trend
in the 220 - 30 K region.}
\end{figure}

Resistivity measurements on single crystal samples of [TbPc$_{2}$]$^0$, with current along the c-axis, have been
performed in the 2.8 - 294 K temperature range in DC mode by using a four-terminal technique. Contacts were made
by attaching  0.01 mm diameter gold wires to the samples with conductive paste and sub-$\mu$A currents were used.
The samples were left free-standing to minimize thermal stress and consequent cracking. The voltage has been
measured with a voltmeter with an internal impedance larger than 10 G$\Omega$. Three different crystals have been
measured and they all showed an increasing resistance on cooling from 2.5 kohm to greater than 2 Mohms below 15 K.
In Fig. (\ref{ro}) the temperature dependence of the resistivity for a crystal with a $2.72\cdot 10^{-3}$
area/lenght ratio is shown. The sample conducts at room temperature with a resistivity $\rho \simeq 6.4$
$\Omega\cdot cm$. Upon cooling, $\rho$ presents an initial smooth decrease till about 220 K and then an activated
growth, characterized by an activation energy $\Delta E \simeq$ $11 meV$, down to about 25 K.

\section{Discussion and Conclusions}

First we shall discuss the temperature dependence of the static
uniform susceptibility $\chi_S$ in [DyPc$_2]^0$, in order to
derive the CF levels structure. Since we are dealing with a
powder, the molecules and the CF axis are randomly oriented with
respect to the external field {\bf {H}} $||$ ${\bf {\hat{z}}}$.
Thus the total magnetization {\bf {M}} can be deduced by summing
the contributions of the molecules with the anisotropy axis
perpendicular and parallel to ${\bf {\hat{z}}}$, namely $M(T)= {2
\over 3}M_{x,y}(T) + {1\over 3}M_z(T)$, with
\begin{equation}
 \label{mxyz}
  M_i(T) = N_A{\sum^{+15/2}_{k=-15/2}\langle\mu^{k}_i\rangle\,e^{-{E^{k}_i\over T}}
            \over
              \sum^{+15/2}_{k=-15/2}e^{-{E^{k}_i\over T}}}
  \;\;\;\; , \;\;\;\; i = x, y, z \;\;\;\; .\vspace{2.5mm}
\end{equation}

In Eq. (\ref{mxyz}), E$^{k}_i$ represents the k-th eigenvalue of the
hamiltonian:
\begin{equation}
 \label{ham}
 \hat{\cal H}= \hat{\cal H}_{CF}+g\mu_B{\bf \hat{J}}\cdot{\bf {H}}
  \;\;\;, \vspace{2.5mm}
\end{equation}
where $\hat{\cal H}_{CF}$ is the crystal field hamiltonian and $g\mu_B {\bf \hat{J}}\cdot{\bf {H}}$ is the Zeeman
term. $\langle \mu^{k}_i\rangle$ is the expectation value of the i-th component of the magnetic moment over the
k-th eigenstate of the hamiltonian. In order to analyze $\chi_S$T we started from the CF structure for the J =
15/2 ground-state multiplet initially derived by Ishikawa {\it et al.} for
[DyPc$_{2}$]$^-\cdot$TBA$^+$.\cite{Ishi5} Then we have varied the splitting among the levels until we found the
best fit of the experimental data. Two sets of possible solutions were found to fit reasonably well $\chi$(T)T
data. However, only one of them yielded a splitting between the lowest lying energy levels around 65 K, the value
derived from AC susceptibility measurements (see later on). The corresponding energy splittings are $\Delta_1$ =
65 K, corresponding to the separation between the $|m = \pm 13/2\rangle $ ground states and the $|m = \pm
9/2\rangle $ first excited levels, $\Delta_2$ = 47 K between the $|m = \pm 9/2\rangle $ levels and the $|m = \pm
11/2 \rangle $ second excited levels and $\Delta_3$ = 460 K between the $|m = \pm 11/2\rangle $ levels and the $|m
= \pm 15/2\rangle $ third excited levels. These values differ from the ones deduced for [DyPc$_{2}$]$^-$ on the
basis of crystal field calculations ($\Delta_1 \simeq$ 50 K, $\Delta_2 \simeq$ 207 K, $\Delta_3 \simeq$ 126 K)
\cite{Ishi5}. As it can be seen in Fig. (\ref {chiT}), a good fit of the experimental data is found.

\begin{figure}[h]
\includegraphics[width=20
pc]{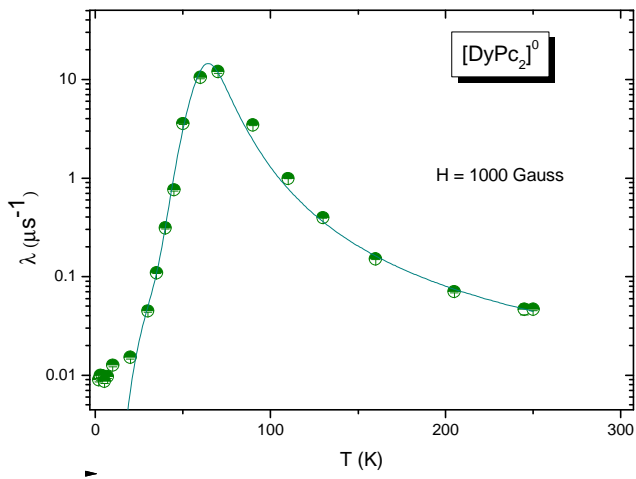}\hspace{0.6pc}%
\caption{\label{picchiMSR}Temperature dependence of the muon longitudinal relaxation rate in [DyPc$_{2}$]$^0$ for
H = 1000 Gauss (circles). The line is the best fit according to Eqs. (\ref{lambdaeq}, \ref{taum}).}
\end{figure}

From the CF splittings, combined with the $\mu$SR relaxation data (Fig. (\ref{picchiMSR})), it is possible to
derive information on the spin-phonon coupling driving the high T spin fluctuations in [DyPc$_{2}$]$^0$. Since the
energy difference between the muon hyperfine levels and the {\it m} levels of Dy$^{3+}$ spin is large, the muon
longitudinal relaxation rate $\lambda$ is driven by an indirect relaxation mechanism involving a muon spin flip
without change in {\it m}. This is possible thanks to the tensorial nature of the hyperfine coupling constant,
which allows the coupling of the transverse components of the hyperfine field ${\it h_{x,y}}$ to $\it J_z$. Thus,
denoting with ($\tau_m$) the finite life-time of the crystal field levels induced by the spin-phonon scattering
processes, $\lambda$ can be written in the form\cite{Lasc}
\begin{equation}
\vspace{2.5mm} \label{lambdaeq}
      {\lambda}={{\gamma^{2}_{\mu}\langle\Delta h^{2}_\bot\rangle}\over {Z}}\,\sum^{+15/2}_{m=-15/2}{\tau_me^{-E_m/T}\,\over
1+\omega^{2}_L \tau^{2}_m}
  \;\;\;\;\;,\vspace{2.5mm}
\end{equation}
$E_m$ being the eigenvalues of the CF levels and $Z$ is the corresponding partition function. It is noted that the
low magnetic field (1000 Gauss) applied during $\mu$SR experiments yields a negligible correction to $E_m$ and,
hence, its effect is negible for $k_BT\gg \mu_BH$. The life-time for the {\it m} levels can be expressed in terms
of the transition probabilities ${\it p_{m,m \pm 1}}$ between {\it m} and ${\it m\pm 1}$ levels, which depend on
the CF eigenvalues and on the spin-phonon coupling constant C \cite{Villain}:
\begin{equation}
\vspace{2.5mm} \label{taum}
      {1\over \tau_m}=p_{m,m-1}+ p_{m,m+1}
  \;\;\;,
\end{equation}
\begin{equation}
 \label{taum}
  p_{m , m \pm 1}=
  C
 {
   {(E_{m \pm 1}-E_m)^{3}}
   \over
   {e^{(E_{m \pm 1}-E_m)/T} -1}
 }
 \;\;\;\;
\vspace{2.5mm}
\end{equation}
%

Now the temperature dependent muon relaxation rate can be correctly fit using Eqs.\ref{lambdaeq}-\ref{taum} (Fig.
\ref{picchiMSR}) by considering the CF level splitting previously estimated from $\chi_S$T fit and by using three
different spin-phonon constants C$_1$ and C$_2$ $\rightarrow 0$, while C$_3 \simeq 2000$ Hz/K$^3$, which should be
associated with the transitions $|{\it m}=\pm 11/2\rangle \leftrightarrow |{\it m}=\pm 13/2\rangle$, $|{\it m}=\pm
9/2\rangle \leftrightarrow |{\it m}=\pm 11/2\rangle$ and $|{\it m}=\pm 15/2\rangle \leftrightarrow |{\it m}=\pm
13/2\rangle$, respectively. We remark that C increases with the energy jump involved in the transition. This
suggests that the relevant processes driving the transitions preferentially involve high energy vibrational modes.

In Fig.(\ref{picchiMSR}) one can notice that the best fit according to Eq.\ref{lambdaeq} reproduces the muon
relaxation data very well for T $>$ 25 K. Since in this model only the thermally activated spin excitations are
concerned, the behaviour of $\lambda$(T) below 25 K should be probably ascribed to different processes. For
instance also in $\chi_S$T a discrepancy from the theoretical calculation is observed at very low T. In fact, for
T $<$ 4 K $\chi_S$T abruptly upturns, possibly signaling the onset of intermolecular correlations. In order to
clarify this point, $\mu$SR experiments have been performed down to very low T in [DyPc$_{2}$]$^0$ and
[TbPc$_{2}$]$^0$ compounds, as shown in Fig. 6. As noted earlier, while the muon relaxation rate is nearly
T-independent below 1 K for both complexes (at least down to 50 mK), the very weak maxima located at $T=0.2$ K and
$0.7$ K for the Dy and Tb samples, respectively, may in fact signal an onset of these intermolecular correlations.

\begin{figure}[h!]
\vspace{11cm} \includegraphics{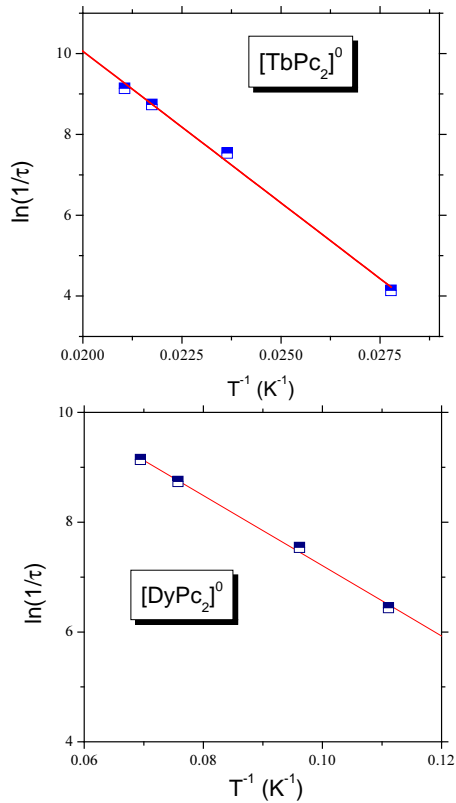}\caption{\label{lntau}Natural
logarithm of the spin correlation rate against the inverse of the $\chi''/ \chi_S$ peak temperature for
[TbPc$_{2}$]$^0$ ($H= 1000$ Gauss) (top) and for [DyPc$_{2}$]$^0$ ($H= 9000$ Gauss) (bottom). The red lines
correspond to the best fits for an activated behaviour for a barrier $\Delta= 750$ K for [TbPc$_{2}$]$^0$ and
$\Delta= 65$ K for [DyPc$_{2}$]$^0$. }
\end{figure}

Now we turn to the comparison of the $\mu$SR and AC susceptibility results. The maxima in $\chi''/\chi_S$ shifting
to lower temperature upon decreasing the irradiation frequency clearly indicate a progressive slowing down of the
dynamics on cooling. In the case of a monodispersive dynamical relaxation mode, at the peak temperature ($T_m$)
the bulk magnetization relaxation time $\tau_c(T_m)$ matches the inverse of the angular frequency $\omega$ of the
applied oscillating field, according to the expression
\begin{equation}
\vspace{2.5mm} \label{chi2}
      \chi''(\omega) = {\chi_S\omega\tau_c \over 1 + \omega^2\tau_c^2}
  \;\;\;,
\end{equation}
We have checked the validity of this expression by performing frequency scans at a fixed temperature and found
that in the explored T range the above expression was satisfied. The linear relation of $ln(\tau_c(T_m)^{-1})$ to
$1/T$ (Fig. \ref{lntau}) indicates that the Orbach process is dominant in the high temperature range and that the
Arrhenius law $\tau_c= \tau_0exp(\Delta/T)$ for the correlation time is obeyed. From the fit of [TbPc$_2$]$^0$
data in Fig.\ref{lntau} one estimates a value for the energy barrier $\Delta \simeq 750 K$, slightly smaller than
the one deduced from muon relaxation rate measurements \cite{Neutri} in the same sample. On the other hand it
should be remarked that this value is much larger than the one reported by Ishikawa et al. ($\Delta\simeq 590$ K)
in the same nominal compound.\cite{Ishi2} The fit of [DyPc$_2$]$^0$ data in Fig.\ref{lntau} gives a much smaller
activation energy $\Delta\simeq 65$ K, which corresponds quite well to the one obtained from the analysis of the
static susceptibility data in Fig.(\ref{chiT}), and provides an estimate for the separation among the lowest
energy levels of [DyPc$_2$]$^0$. This value is much different with the barrier estimated from $\mu$SR relaxation
at higher temperature, which yields an estimate for the energy separation among the high energy levels.
\cite{Neutri}

\begin{figure}[h!]
\vspace{6.8cm} \includegraphics{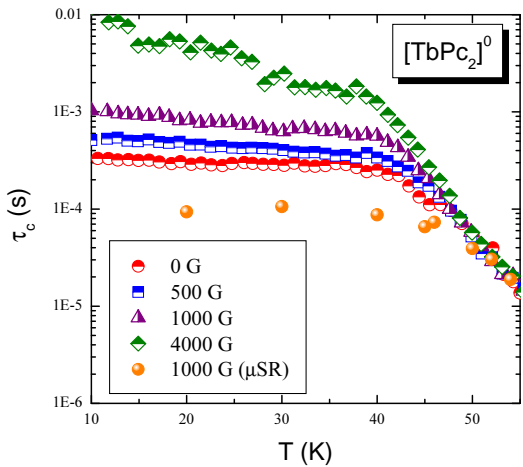}
\caption{\label{chiACvsH}Temperature dependence of $\tau_c$, deduced from Eq.(\ref{chi2}) for a static field of 0,
500, 1000, 4000 Gauss in [TbPc$_{2}$]$^0$ compound. Solid circles correspond to $\tau_c$ temperature dependence as
derived from $\mu$SR analysis.}
\end{figure}

In order to better compare the T-dependence of the correlation time derived by means of AC susceptibility and
$\mu$SR in [TbPc$_2$]$^0$ we have derived, via Eq.\ref{chi2}, the behaviour of $\tau_c$ estimated from AC
susceptibility measurements and compared to the data reported in Fig. (\ref{tau}), derived with $\mu$SR. As it is
shown in Fig.(\ref{chiACvsH}) the T-dependence of $\tau_c$ derived by the two techniques at different magnetic
fields overlap rather well above 45 K. At lower temperatures a plateau is evidenced by both techniques and
associated with tunneling processes among the $|m=\pm 6\rangle$ low-energy levels. It is noticed that the plateau
lies at different values depending on the magnitude of the applied field and on the technique. As it has been
pointed out in Ref.\onlinecite{Neutri} the magnetic field leads to a Zeeman splitting of the two-fold degenerate
ground-state and causes a reduction of the tunneling rate. Although this justifies the progressive increase of
$\tau_c$ with magnetic field observed by each technique, it cannot explain the significant difference in the
$\tau_c$ values deduced by $\mu$SR and AC susceptibility in the low-temperature tunneling regime. Thus, while both
techniques probe the same high temperature activated dynamics driven by spin-phonon coupling, the low-temperature
tunneling dynamics probed at the microscopic or at the macroscopic level are different. This suggests that there
is some correlation among the magnetic moments in the different molecular units which cause fluctuations which do
not affect the total magnetization. This coupling should lead to flip-flop like fluctuations \cite{Abragam}
yielding a magnetic moment flip from $|m=+6\rangle$ to $|m=-6\rangle$ on one molecule and the opposite flip in the
adjacent molecule. These processes, which cannot be accounted for by a direct dipolar coupling among Tb$^{3+}$
spins since they produce a change $\Delta m= \pm 12$, do not yield a net variation in the total magnetization and,
hence, do not contribute to the AC susceptibility. In other words, at low-T $\mu$SR probes $T_2$-like processes
involving the fluctuations of Tb$^{3+}$ moments while AC susceptibility probes only those $T_1$-like processes
yielding a net variation of the macroscopic magnetization. The precise nature of these fluctuations, similar to
the ones involving phonon trapping in Ni10,\cite{Ni10} still has to be clarified.

Finally, we turn to the discussion of transport properties. As it is shown in Fig.(\ref{ro}), the temperature
dependence of the resistivity is characterized by a rather small energy barrier, around 11 meV. Now, in
[TbPc$_2$]$^0$ molecular orbital calculations \cite{Huckel} show that there is one unpaired electron in the $a_2$
highest occupied molecular orbital (HOMO), basically involving orbitals from the carbon atoms at the center of the
Pc rings. If there is a sufficient overlap between the $a_2$ orbitals of the adjacent molecules, which are stacked
forming chains, electron delocalization and a metallic behaviour should be attained. However, in those Pc-based
systems the electron correlations are significant, particularly at half-band filling \cite{Filibian}, and the
on-site Coulomb repulsion $U$ can overcome the hopping integral $t$ among adjacent molecules leading to a Hubbard
insulating behaviour. Hence, the activated behaviour of the resistivity should be ascribed to correlation effects
rather than to the band structure, which should typically lead to much larger activation energies. The band
structure and, accordingly, the hopping integral $t$ appear to sizeably depend on the local structure, in
particular, on the buckling of the Pc ring plane and on the rotation of adjacent Pc rings, hence it is possible
that at high T the crossover from a negative  to a positive $d\rho/dT$ with increasing T, might be ascribed to
those structural effects.

In conclusion, we have clearly shown that both AC susceptibility and $\mu$SR probe a high T activated spin
dynamics in neutral [LnPc$_2$] molecular magnets, characterized by quantitatively similar correlation times. At
low temperature both techniques exhibit a flattening of the correlation time which is associated with tunneling
processes. Nevertheless, from $\mu$SR one estimates correlation times which are one order of magnitude shorter
than the ones derived by AC susceptibility measurements. This discrepancy between microscopic and macroscopic
techniques indicates that there must be some correlation among Tb$^{3+}$ moments, which cause fluctuations which
do not contribute to the uniform macroscopic susceptibility. The analysis of the static uniform susceptibility and
of the $\mu$SR relaxation rates in [DyPc$_2$]$^0$ have allowed for a better characterization of the CF splitting
of the $J=15/2$ multiplet and of the spin-phonon couplings driving the activated dynamics in this compound than
was reported in Ref.\onlinecite{Ishi5}. Finally, we have reported the behaviour of the resistivity in
[TbPc$_2$]$^0$, where a low T activated behaviour, possibly arising from electron-electron Hubbard-like
correlations, is evidenced.

\section*{Acknowledgements}

Technical support from C. Baines at PSI and from C. Dhital at Boston College is gratefully acknowledged. The
research activity in Pavia was supported by Fondazione Cariplo (Grant N. 2008-2229) research funds, the activity
in Karlsruhe by the ERA-Chemistry project "MULTIFUN". We would like to thank Dr. O. Fuhr for carrying out the
X-Ray diffraction analysis. Work at Boston College was supported by National Science Foundation grant No.
DMR-0710525.



\end{document}